\documentclass[twocolumn,amsmath,amssymb,aps,prb,floatfix,10pt,showpacs]{revtex4-1}
\usepackage[utf8]{inputenc}
\usepackage{graphicx}
\usepackage{dcolumn}
\usepackage{bm}
\usepackage{booktabs}
\usepackage{multirow}
\usepackage{array}
\usepackage{mathrsfs}
\usepackage{braket}
\usepackage{mathtools}
\usepackage{subfigure}
\usepackage{hyperref}
\usepackage{amsmath}
\usepackage{nicefrac}
\usepackage[usenames,dvipsnames]{xcolor}

\hypersetup{pdfauthor={Benjamin Siegert, Andrea Donarini, and Milena Grifoni},pdftitle={Effects of spin-orbit coupling and many-body correlations in STM transport through copper phthalocyanine}}
 
\allowdisplaybreaks

\newcommand{\erz}[2]{\hat{\operatorname{#1}}_{#2}^{\dag}}			
\newcommand{\ver}[2]{\hat{\operatorname{#1}}_{#2}^{\phantom{\dag}}}		
\newcommand{\rr}{\mathbf{r}}							
\newcommand{\kk}{\mathbf{k}}							
\newcommand{\HH}[1]{\hat{\operatorname{H}}_{\mathrm{#1}}}		  	
\newcommand{\VV}[1]{\hat{\operatorname{V}}_{\mathrm{#1}}}			
\newcommand{\eps}[1]{\epsilon_{#1}}						
\newcommand{\oost}{\frac{1}{\sqrt{2}}}						
\newcommand{\up}{\uparrow}							
\newcommand{\down}{\downarrow}	
\newcommand{\ee}[1]{\mathrm{e}^{#1}}						

\begin{document}

\title{Effects of spin-orbit coupling and many-body correlations in STM transport through copper phthalocyanine}
\author{Benjamin Siegert}
\email{benjamin.siegert@ur.de}

\author{Andrea Donarini}
\author{Milena Grifoni}
\affiliation{Institut f\"ur Theoretische Physik, Universit\"at Regensburg, D-93040 Regensburg, Germany}

\date{\today}

\begin{abstract}
The interplay of exchange correlations and spin-orbit interaction (SOI) on the many-body spectrum of a copper phtalocyanine (CuPc) molecule and their
signatures in transport are investigated. We first derive a minimal model Hamiltonian in a basis of frontier orbitals which is able to reproduce experimentally observed singlet-triplet splittings;
in a second step SOI effects are included perturbatively. Major consequences of the SOI are the splitting of former degenerate levels and a magnetic anisotropy, 
which can be captured by an effective low-energy spin Hamiltonian.
We show that STM-based magnetoconductance measurements can yield clear signatures of both these SOI induced effects.
\end{abstract}

\maketitle

\section{Introduction}

Spin-orbit interaction (SOI) can play a major role in molecular spintronics. 
For example, in combination with the configuration of the non-magnetic component (organic ligand), it is known to be 
 essential in establishing magnetic anisotropy in high-spin molecular magnets~\cite{Gatteschi2006}. 
 Effective spin-Hamiltonians are commonly used to describe this anisotropy, and usually well capture the low energy properties of these systems, 
 see e.g. Ref.~\cite{Mannini2010}. Such effective Hamiltonians have been derived microscopically  for widely studied molecular magnets like 
Fe$_8$, Fe$_4$ and Mn$_{12}$~\cite{Chiesa2013}. Recently, magnetic anisotropy effects could be directly probed by magnetotransport spectroscopy for 
Fe$_4$ in  quantum dot setups~\cite{Wegewijs2015,Burzuri2015}. An interesting question is hence if other classes of metallorganic compounds, 
like the widely studied metal phthalocyanines~\cite{params:liao,mugarza2012}, exhibit magnetic anisotropy induced by the interplay of electronic correlations and SOI. 
Indeed, in an XMCD analysis copper phthalocyanine (CuPc) was found to exhibit enormous anisotropies in both spin and orbital
dipole moments~\cite{stepanow2010}.
Furthermore, recent experimental findings for cobalt pththalocyanine in an STM setup~\cite{nature_STM} suggest that many-body correlations  play an
important role in the interpretation of the transport measurements. 
In recent work~\cite{sco_cupc_arxiv}, we have explictly investigated long range and short 
range electron-electron correlations effects in CuPc
and found a singlet-triplet splitting of the former anionic groundstate of about 18 meV, and thus a triplet as anionic ground state. 

In this work we add the SOI to our analysis. We find that it further removes the triplet degeneracy by inducing splittings of few tenths of meV. 
Moreover, in combination with exchange correlations, it produces a 
 magnetic anisotropy which can in turn be captured by an effective spin Hamiltonian.
 
In general, the accurate calculation of the many-body properties of metallorganic molecules, like the molecular magnets or our CuPc, is a highly nontrivial task. 
In fact, the number of their atomic constituents is large enough that
exact diagonalization is not possible and standard density-functional schemes have difficulties in capturing short ranged electron-electron correlations~\cite{Chiesa2013}. 
 In order to reduce the size of the many-body Fock space, we use a basis of frontier molecular orbitals as the starting point to include electronic correlations~\cite{ryndyk2013,sco_cupc_arxiv}
and construct a generalized Hubbard Hamiltonian. Furthermore, the symmetry of the molecule greatly helps to reduce the number of matrix elements one has to calculate in this basis. 

To probe both SOI induced splittings and magnetic anisotropy, we further investigated the current characteristics of a CuPc molecule in an STM configuration 
similar to the experiments in Refs.~\cite{repp_meyer_2005,repp_charge_state}: the molecule is put on a thin insulating layer grown on top of a conducting substrate. 
The layer functions as a tunneling barrier and decouples the molecule from the substrate. Hence the CuPc molecule acts as a molecular quantum dot weakly coupled 
by tunneling barriers to metallic leads (here the STM tip and the substrate).
This quantum dot configuration should be favourable to experimentally probe SOI splittings and magnetic anisotropies when an external magnetic field is applied to the system, 
in analogy to the experiments in Ref.~\cite{Burzuri2015}. Indeed, we demonstrate that experimentally resolvable SOI splitting should be observed at magnetic fields of a few Tesla. 

The paper is organized as follows: In Sec.~\ref{Sec2} we derive a microscopic Hamiltonian for CuPc in the frontier orbital basis which includes exchange correlations and the SOI. 
This Hamiltonian is diagonalized exactly and used in further spectral analysis and transport calculations. Its spectrum is also used to benchmark the prediction of an 
effective spin Hamiltonian which well captures the low energy properties of CuPc both in its neutral and anionic configurations. Finally, transport calculations with and 
without magnetic fields are presented and SOI induced signatures are analyzed.
Section~\ref{Sec3} contains our conclusions.

\begin{figure}
 \includegraphics[width=\columnwidth]{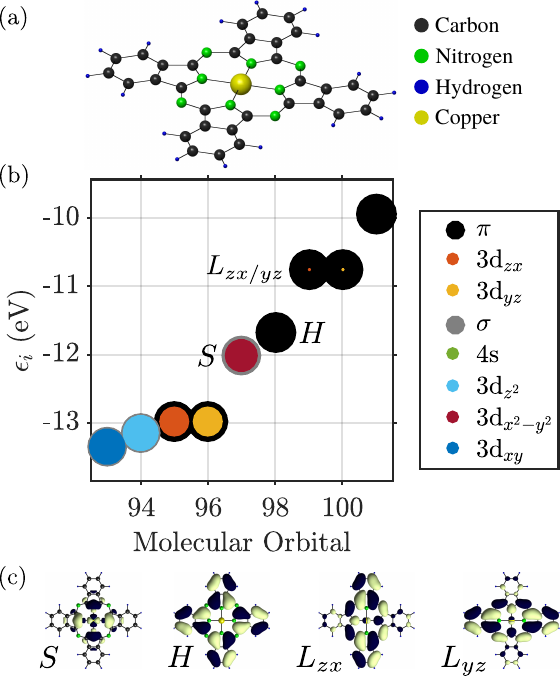}
 \caption{ (a) Geometry and atomic composition of CuPc. (b) Single particle energies of relevant molecular orbitals.
		Black (grey) circles depict the $\pi$ ($\sigma$) character of the corresponding orbital. %
	The color (diameter) of the inner circles characterizes the type (weight) of the metal orbital contribution on the corresponding molecular orbital.
 (c) Depiction of the four frontier orbitals retained in this work: SOMO ($S$), HOMO ($H$) and LUMO$_{zx/yz}$ ($L_{zx/yz}$).}\label{fig_1}
\end{figure}

\section{Results and Discussion}\label{Sec2}
\subsection{Microscopic model Hamiltonian for CuPc}
The focus of this section is the establishment of a minimal model Hamiltonian for an isolated CuPc molecule capable to account for both electron-electron interaction
and spin-orbit coupling effects. As discussed below, parameters are fixed such that experimental observations for the singlet-triplet splitting~\cite{mugarza2012} as well as positions
of anionic and cationic resonances~\cite{repp_charge_state} are satisfactorily reproduced.
In its most general form and for a generic molecule such Hamiltonian reads
%
 \begin{align}\label{Eq:H_mol}
	\HH{mol}  =  \HH{0} + \VV{ee}  + \VV{SO},
 \end{align}
where the single-particle Hamiltonian of the molecule is given by $\HH{0}$, $\VV{ee}$ describes electronic interactions and $\VV{SO}$ accounts for the spin-orbit interaction (SOI).
\subsubsection{Single particle Hamiltonian for CuPc}

The one-body Hamiltonian $\HH{0}$, written in the atomic basis $\ket{\alpha}$, reads
%
\begin{align}\label{Eq:H_onebody}
 \HH{0}=\sum_{\alpha\beta\sigma}\left( \eps{\alpha}\delta_{\alpha\beta} + b_{\alpha\beta} \right)\erz{d}{\alpha\sigma}\ver{d}{\beta\sigma},
\end{align}
where $\alpha$ is a multi-index combining atomic species and orbital quantum number at position $\rr_\alpha$, see Fig.~\ref{fig_1} (a).
For the ligand we consider the set of all 2s (1s for hydrogen),
2p$_x$ and 2p$_y$ orbitals as the $\sigma$-system, and consequently the set of 2p$_z$ orbitals as 
the $\pi$-system. On the metal, the 3d$_{xy}$, 3d$_{x^2-y^2}$, 3d$_{z^2}$ and 4s orbitals contribute to the $\sigma$-system, while
the 3d$_{zx}$ and 3d$_{yz}$ belong to the $\pi$-system.
This basis yields a total of 195 valence electrons for neutral CuPc.
Atomic onsite energies $\eps{\alpha}$ and geometrical parameters were taken from Refs.~\cite{mann_hf,params:liao}.
The hopping matrix elements $b_{\alpha\beta}$ in Eq.~\eqref{Eq:H_onebody} are obtained by using the Slater-Koster\cite{slaterlcao} and Harrison\cite{harrisonlcao} LCAO schemes, 
similar to Ref.~\cite{h2pcpaper}.
Numerical diagonalization of $\HH{0}$ finally yields single particle energies $\eps{i}$, see Fig.~\ref{fig_1} (b),
and molecular orbitals $\ket{i\sigma}=\sum_\alpha c_{i\alpha}\ket{\alpha\sigma}$, cf. App.~\ref{App:AO2MO}.

Stemming from Hartree-Fock calculations for isolated atoms~\cite{mann_hf}, the atomic onsite energies $\eps{\alpha}$ do not take into account
the ionic background of the molecule and crystal field contributions. Therefore, molecular orbital energies $\eps{i}$
have to be renormalized with parameters $\delta_i$ to counteract this shortage, yielding (cf. App.~\ref{Sec:reduction})
\begin{align}\label{Eq:renorm}
 \HH{0} = \sum_{i\sigma} \left( \eps{i} + \delta_i \right) \hat{n}_{i\sigma}.
\end{align}
In this work we use a constant shift $\delta_i=\delta=1.5~$eV.

Due to the odd number of valence electrons, in its neutral configuration CuPc has a singly occupied molecular orbital (SOMO). The latter also does not become doubly occupied when the molecule is 
in its anionic groundstate~\cite{params:liao}. Hence, the orbitals most relevant for transport (frontier orbitals)
are the SOMO ($S$), the HOMO ($H$) and the two degenerate LUMOs ($L_{zx/yz}$), which transform according to the $b_{1g}$, $a_{1u}$ and $e_g$ irreducible 
representations of the point group of CuPc (D$_{4h}$), respectively.
They are depicted in Fig.~\ref{fig_1} (c).
The LUMO orbitals in their real-valued representations, $\ket{L_{zx}}$ and $\ket{L_{yz}}$, have equal contributions $c_{L}\approx0.097$ on either 3d$_{zx}$ and 3d$_{yz}$ orbitals on the metal, respectively. 
Due to their degeneracy, they can be transformed into their complex, rotational invariant representations:
\begin{align}\label{Eq:lumos}
 \ket{L\pm} 	=& \mp 	2^{-\nicefrac{1}{2}}\, 			\Big( 	\ket{L_{zx}} 				\pm i	\ket{L_{yz}} 				\Big)\nonumber\\
		=& \mp 	2^{-\nicefrac{1}{2}}\, 	\sqrt{1-c_L^2}\,\Big( 	\ket{L_{zx}}_\mathrm{Pc} 		\pm i	\ket{L_{yz}}_\mathrm{Pc} 		\Big)\nonumber\\
		& \mp  	2^{-\nicefrac{1}{2}}\,	c_L		\Big(  	\ket{3\mathrm{d}_{zx}}_\mathrm{Cu}  	\pm i 	\ket{3\mathrm{d}_{yz}}_\mathrm{Cu}  	\Big) \nonumber\\
		=& \sqrt{1-c_L^2}\,\ket{L\pm}_\mathrm{Pc} +c_L \ket{3,2,\pm1}_\mathrm{Cu},
\end{align}
where $\ket{3,2,\pm1}_\mathrm{Cu}$ is the $n=3$ metal orbital with angular momentum $\ell=2$ and magnetic quantum number $m=\pm1$.
To distinguish contributions from the pure phthalocyanine (Pc) ligand  and the copper (Cu) center, 
we introduced $\ket{\cdot}_\mathrm{Pc}$ and $\ket{\cdot}_\mathrm{Cu}$, respectively.
Likewise, with $c_S\approx0.90$, we can write for the SOMO:
\begin{align}
 \ket{S} 	&= \sqrt{1-c_S^2}\,\ket{S}_\mathrm{Pc} + c_S\ket{3\mathrm{d}_{x^2-y^2}}_\mathrm{Cu}\nonumber\\
		&= \sqrt{1-c_S^2}\,\ket{S}_\mathrm{Pc} + 2^{-\nicefrac{1}{2}}\,c_S \Big( \ket{3,2,-2}_\mathrm{Cu} + \ket{3,2,2}_\mathrm{Cu} \Big),
\end{align}
where $\ket{3,2,\pm2}_\mathrm{Cu}$ is the $n=3$ metal orbital with angular momentum $\ell=2$ and projection $m=\pm2$ onto the $z$-axis.
Finally, the HOMO has no metal contributions and thus we have trivially $\ket{H}=\ket{H}_\mathrm{Pc}$.
The representations introduced in Eq.~\eqref{Eq:lumos} have the advantage that the four frontier orbitals can then be characterized by the phases $\varphi_i$ acquired under rotations of $\frac{\pi}{2}$
around the main molecular symmetry axis; for the SOMO $\varphi_S=\pi$, for the HOMO $\varphi_H=0$ and for the two LUMOS $\varphi_{L\pm}=\pm\frac{\pi}{2}$.

\subsubsection{Many-body Hamiltonian in the frontier orbitals basis}

In order to set up a minimal many-body Hamiltonian, we restrict the full Fock space 
to many-body states spanned by the SOMO ($S$), the HOMO ($H$) and the two LUMO ($L\pm$) orbitals and write Eq.~\eqref{Eq:H_mol} in this basis.
Hence, for neutral CuPc the number of electrons populating the frontier orbitals is $N_0=3$.

We exploit the distinct phases acquired by the frontier orbitals under 90 degree rotations to determine selection rules for the matrix elements $V_{ijkl}$ in $\VV{ee}$,
\begin{align}\label{Eq:V_ee_gen}
 \VV{ee} &= \sum_{ijkl}\sum_{\sigma\sigma'}\,V_{ijkl}\,\erz{d}{i\sigma}\erz{d}{k\sigma'}\ver{d}{l\sigma'}\ver{d}{j\sigma},
\end{align}
namely $V_{ijkl}\neq0$ if $\phi_i-\phi_j+\phi_k-\phi_l=2\pi\cdot n,~ n\in\mathbb{Z}$, cf. App.~\ref{Sec:reduction}.
Equation~\eqref{Eq:V_ee_gen} in this basis then reads
\begin{widetext}
 \begin{align}\label{Eq:V_ee}
	\VV{ee}  &=  \sum_iU_{i}\,\hat n_{i\uparrow}\hat n_{i\downarrow}  
	+\frac{1}{2}\sum_{[ij]}  U_{ij}\,\hat n_i \hat n_j 																
	-\frac{1}{2}\sum_{[ij]} \sum_\sigma J_{ij}^\mathrm{ex}\,\left(  \hat n_{i\sigma} \hat n_{j\sigma}  -  \erz{d}{i\sigma}\erz{d}{j\bar\sigma}\ver{d}{i\bar\sigma}\ver{d}{j\sigma}  \right)   \nonumber\\ 		
	&+\frac{1}{2}\sum_{[ij]}  \sum_\sigma J_{ij}^\mathrm{p}\,\erz{d}{i\sigma}\erz{d}{i\bar\sigma}\ver{d}{j\bar\sigma}\ver{d}{j\sigma}  
	+\frac{1}{2}\sum_{[ijk]}  \sum_\sigma \left(  \tilde J_{ijk}^\mathrm{p}\,\erz{d}{i\sigma}\erz{d}{i\bar\sigma}\ver{d}{k\bar\sigma}\ver{d}{j\sigma}+\mathrm{h.c.} \right) 	
 \end{align}
  \end{widetext}
where the indices $i,j,k,l$ now run over the set of frontier orbitals, and the notation $[ijkl]$ means that the sum runs only over unlike indices, i.e. $i$, $j$, $k$, and $l$
 are different from each other in the corresponding sum. 
The abbreviations we introduced in Eq.~\eqref{Eq:V_ee} are the orbital Coulomb interaction $U_i=V_{iiii}$, the
 inter-orbital Coulomb interaction $U_{ij}=V_{iijj}$, the exchange integral $J_{ij}^\mathrm{ex}=V_{ijji}$, the ordinary pair hopping term $J_{ij}^\mathrm{p}=V_{ijij}$ and the split pair hopping term 
 $\tilde J_{ijk}^\mathrm{p}=V_{ijik}$. 
 Contributions with four different indices are found to be very small (on the order of $\mu$eV) and thus omitted in this work.
 The matrix elements $V_{ijkl}$ are calculated numerically using Monte Carlo
integration\cite{gsl} and renormalized with a dielectric constant $\varepsilon_{r}=2.2$ in order to account for screening by frozen orbitals~\cite{ryndyk2013}. A table (cf. Tab.~\ref{tab_1}) with
the numerically evaluated interaction constants is found in App~\ref{Sec:reduction}.
 
\subsubsection{Spin-orbit interaction (SOI) in the frontier orbitals basis}
 
A perturbative contribution  to the bare one-body Hamiltonian $\HH{0}$ relevant in molecular systems is provided by the SOI. 
In the following we derive an effective spin-orbit coupling operator acting on the subset of frontier orbitals. 
The atomic SOI operator reads
\begin{align}\label{Eq:SO1}
 \VV{SO} = \sum_{\alpha,\ell_\alpha} \xi_{\ell_\alpha}\, \hat{{\bm{\ell}}}_\alpha\cdot\hat{{\bm{s}}}_\alpha ,
\end{align}
where $\alpha$ and $\ell_\alpha$ run over all atoms and shells, respectively. By evaluating Eq.~(\ref{Eq:SO1}) only on the central copper atom, i.e. $\ell=2$ and $\alpha=\mathrm{Cu}$,
$\VV{SO}$ in second quantization is given by
\begin{align}\label{Eq:SO_Cu}
		\VV{SO} = \xi_\mathrm{Cu} \Bigg( \sum_{m=-2}^{2} 
\frac{m}{2}\left( \erz{d}{m\uparrow} \ver{d}{m\uparrow} - \erz{d}{m\downarrow} \ver{d}{m\downarrow} \right) \notag\\
+\sqrt{\frac{3}{2}}	
\left(	\erz{d}{0 \downarrow} \ver{d}{-1 \uparrow} + \erz{d}{1 \downarrow} \ver{d}{0 \uparrow} + \mathrm{h.c.} \right ) \notag\\
+
\left(	\erz{d}{2 \downarrow} \ver{d}{1 \uparrow} +	\erz{d}{-1 \downarrow} \ver{d}{-2 \uparrow} + \mathrm{h.c.} \right )
\Bigg),
\end{align} 
where $\erz{d}{m\sigma}$ creates an electron with spin $\sigma$ on the copper atom in the orbital specified by $(\ell=2,m)$.
For an electron in the 3d-shell of Cu we use $\xi_\mathrm{Cu}\approx100~$meV~\cite{spin_orbit_params}.
Projecting Eq.~\eqref{Eq:SO_Cu} onto the minimal set of frontier orbitals then yields:
\begin{align}
	\VV{SO} =	&	\lambda_1 \sum_{\tau=\pm} \tau\left( \erz{d}{L\tau\uparrow}\ver{d}{L\tau\uparrow} - \erz{d}{L\tau\downarrow}\ver{d}{L\tau\downarrow} \right) \notag\\
			+& 	\lambda_2\left( \erz{d}{S\uparrow}\ver{d}{L-\downarrow} + \erz{d}{L+\uparrow}\ver{d}{S\downarrow} + \mathrm{h.c.} \right) ,
\end{align} 
where $\lambda_1=\frac{1}{2}\xi_\mathrm{Cu}\,|c_L|^2=0.47~$meV and $\lambda_2=\xi_\mathrm{Cu}\,\frac{c_Sc_L}{\sqrt{2}}=6.16~$meV
are now effective spin-orbit coupling constants. A similar analysis of SOI in CuPc, laying more focus on the central Cu atom, can be found in Ref.~\cite{yu2012}.

Finally, many body eigenenergies $E_{Nk}$ and eigenstates $\ket{Nk}$, labelled after particle number $N$ and state index $k$, are obtained
by exact numerical diagonalization of $\HH{mol}$ in the frontier orbitals basis.
Despite numerically tractable, the problem described by $\HH{mol}$ is still highly intricate, as the Fock space has dimension $4^4=256$.
In reality, though, only few low-lying many-body states are relevant at low energies, what enables further simplification and even an analytical treatment,
as discussed in the next subsection.
 
\subsection{Low-energy spectrum of CuPc and effective spin Hamiltonian}

\begin{figure}
 \includegraphics[width=\columnwidth]{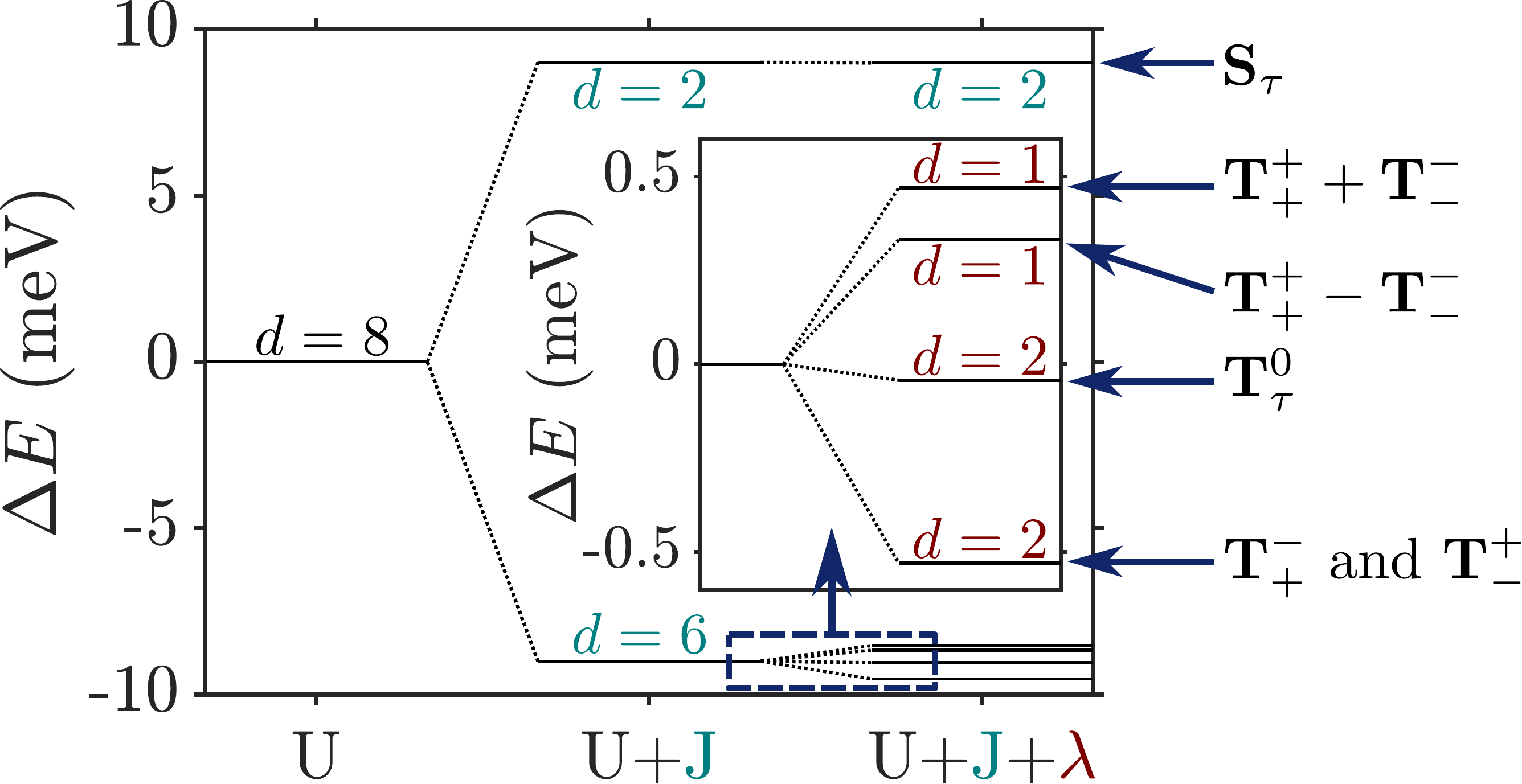}
 \caption{ Lowest lying anionic states of CuPc, together with their grade of degeneracy $d$. 
 Without exchange and SOI, the anionic groundstate is eightfold degenerate. When exchange interaction between SOMO and LUMOs is introduced,
 the degeneracy is lifted, yielding two triplets and two singlets because of the orbital degeneracy of the LUMO. SOI further splits the triplet states, generating a twofold degenerate anionic groundstate
 consisting of the states $\mathbf{T}_+^-$ and $\mathbf{T}_-^+$.}\label{fig_2}
\end{figure}

In the following we will analyze the neutral and anionic low-energy part of the many-body spectrum of CuPc
and establish an effective Hamiltonian which enables us to analyze the low-energy behaviour in a more lucid way.
To this extent, we start by observing that $\HH{mol}$ (in the considered particle number subblocks) contains different energy scales, 
in particular, $U>J>\lambda$, what suggests a hierarchy of steps. We use $U$, $J$ and $\lambda$ to denote the set of all Hubbard-like parameters ($U_i,U_{ij}$), 
all exchange parameters ($J^\mathrm{ex}_{ij}$,$J^\mathrm{p}_{ij}$,$\tilde J^\mathrm{p}_{ijk}$) and all SOI parameters ($\lambda_i$), respectively.
As a first step we set both the exchange ($J$) and SOI ($\lambda$) contributions to $\HH{mol}$ to zero and determine the neutral and anionic groundstates.
In a second and third step exchange and SOI are added, respectively.

\subsubsection{Neutral low-energy spectrum}
In the neutral low-energy part of the spectrum, we retain the two spin-degenerate groundstates of $\HH{mol}(J=0,\lambda=0)$,
\begin{align}\label{neutral_groundstate}
 \ket{N_0,\sigma} := \erz{d}{S\sigma} \ket{\Omega},
\end{align}
with corresponding energy $E_{N_0}^\mathrm{g}$. Here we defined $\ket{\Omega}=\erz{d}{H\up}\erz{d}{H\down}\ket{0}$.
The groundstates in Eq.~\eqref{neutral_groundstate}
are neither affected by $\VV{SO}$ nor by the exchange terms in Eq.~\eqref{Eq:V_ee}. Trivially, the effective Hamiltonian
in the basis of $\ket{N_0,g\sigma}$ reads:
\begin{align}\label{H_eff_neutral}
 \operatorname{H}_{0}^{N_0} = E_{N_0}^\mathrm{g}.
\end{align}
In principle Eq.~\eqref{Eq:V_ee} also contains terms which act on the neutral groundstate, like for example pair hopping terms proportional to 
$\tilde J_{HL+L-}^\mathrm{p}$, and cause admixtures with other many-body states. However, according to our full numerical calculations, 
these admixtures are rather small and do not affect transitions between neutral and anionic states.

\subsubsection{Anionic low-energy spectrum}

Continuing with the anionic low-energy part of the spectrum of $\HH{mol}(J=0,\lambda=0)$, we find an eightfold degenerate groundstate:
\begin{align}
 \ket{N_0+1,\tau\sigma\sigma'} := \erz{d}{S\sigma} \erz{d}{L\tau\sigma'} \ket{\Omega},
\end{align}
with corresponding energy $E_{N_0+1}^\mathrm{g}$.
The eightfold degeneracy comes from the two unpaired spins in either SOMO or LUMO and the orbital degeneracy of the LUMO orbitals.
In order to make the anionic eigenstates also eigenstates of the spin operators $\hat{\mathbf{S}}^2$ and $\hat S_z$, they can be rewritten as
\begin{align}\label{main_states}
	\ket{\mathbf{S}_\tau} 	&= \oost\left(\erz{d}{S\up}\erz{d}{L\tau\down} - \erz{d}{S\down}\erz{d}{L\tau\up}\right)\ket{\Omega},  \nonumber\\
	\ket{\mathbf T_\tau^+}	&= \erz{d}{S\up}\erz{d}{L\tau\up}\ket{\Omega},\nonumber\\
	\ket{\mathbf T_\tau^0} &= \oost\left(\erz{d}{S\up}\erz{d}{L\tau\down} + \erz{d}{S\down}\erz{d}{L\tau\up} \right)\ket{\Omega}, \nonumber\\
	\ket{\mathbf T_\tau^-} &= \erz{d}{S\down}\erz{d}{L\tau\down}\ket{\Omega}.
\end{align}
The orbital degeneracy of the LUMOs, expressed by the index $\tau$, is responsible for the two sets of singlets (total spin $S=0$) and triplets (total spin $S=1$).
Considering exchange interaction in a second step, we find that only the $J_{SL}^\mathrm{ex}$ term in Eq.~\eqref{Eq:V_ee}, 
\begin{align}\label{Eq:V_ee_2}
 -\sum_{\tau\sigma} J_{SL}^\mathrm{ex}\,\left(  \hat n_{S\sigma} \hat n_{L\tau\sigma}  -  \erz{d}{S\sigma}\erz{d}{L\tau\bar\sigma}\ver{d}{S\bar\sigma}\ver{d}{L\tau\sigma}  \right),
\end{align}
directly determines the low-energy structure of the anionic low-energy part because of the singly occupied SOMO and LUMOs: The degeneracy between
singlets and triplets is lifted, see Fig.~\ref{fig_2}, and we obtain
\begin{align}
 E_{\mathbf{S}}=E_{N_0}^\mathrm{g}+J^\mathrm{ex}_{SL},\nonumber\\
 E_{\mathbf{T}}=E_{N_0}^\mathrm{g}-J^\mathrm{ex}_{SL}
\end{align}
for the singlets and triplets, respectively.

Finally, to analyze in a third step how $\VV{SO}$ affects the low-energy part of the anionic part of the spectrum, in particular which degeneracies are lifted,
we treat it as a perturbation and apply second order perturbation theory to obtain the energy shifts.
To this end, some additional states have to be considerd. They are listed in App.~\ref{App:SO1}.

The states $\mathbf{T}^+_-$ and $\mathbf{T}^-_+$ experience a downshift due to $\VV{SO}$ and become the groundstates.
Measuring energies with respect to $E_{\mathbf{T}}$, we get
\begin{align}
 \Delta E_{\mathbf T_+^-} =  \Delta E_{\mathbf T_-^+}	&= - \lambda_1 - \frac{2\lambda_2^2}{\Delta_1+J^\mathrm{ex}_{SL}},
\end{align}
see Fig.~\ref{fig_2}. 
Note that in our numerical calculations $\mathbf T_+^-$ and $\mathbf T_-^+$ are mixed and the degeneracy of the
resulting states is lifted by a small shift in the range of some $\mu$eV. 
A more detailed discussion concerning the mixing of $\mathbf T_+^-$ and $\mathbf T_-^+$ can be found in App.~\ref{App:SO2}.
The next states are $\mathbf{T}_+^0$ and $\mathbf{T}_+^0$ with
\begin{align}
 \Delta E_{\mathbf T_+^0} = \Delta E_{\mathbf T_-^0} = - \frac{\lambda_1^2}{2J^\mathrm{ex}_{SL}}-\frac{\lambda_2^2}{\Delta_1-J^\mathrm{ex}_{SL}}.
\end{align}
Due to their quadratic dependence on $\lambda_1$ and $\lambda_2$, these states change very little with $\VV{SO}$. The degeneracy of the states $\mathbf T_+^+$
and $\mathbf T_-^-$ is lifted by the mixing of these states through $\VV{SO}$. We find
\begin{align}
	\ket{\alpha} &= \oost\Big( \ket{T_+^+} + \ket{T_-^-} \Big), \\
	\ket{\beta} &= \oost\Big( \ket{T_+^+} - \ket{T_-^-} \Big),
\end{align}
where for $\ket{\beta}$ we omitted smaller additional contributions from other states. The energies change according to
\begin{align}
	\Delta E(\alpha) &= \lambda_1, \\
	\Delta E(\beta) &= \lambda_1 - 4 \lambda_2^2\left( \frac{1}{\Delta_1+J^\mathrm{ex}_{SL}} + \frac{1}{\Delta_2+J^\mathrm{ex}_{SL}} \right).
\end{align}
For further details we refer to App.~\ref{App:SO1}.
Finally, the singlets $\mathbf S_+$ and $\mathbf S_-$, similar to $\mathbf T_+^0$ and $\mathbf T_-^0$, change very little (with respect to $E_{\mathbf{S}}$):
\begin{align}
 \Delta E_{\mathbf{S}_\tau} = \frac{\lambda_1^2}{2J^\mathrm{ex}_{SL}}-\frac{\lambda_2^2}{\Delta_1-J^\mathrm{ex}_{SL}}.
\end{align}
By introducing $\hat\tau := \hat n_{L+} - \hat n_{L-}$,
%
%
an approximate Hamiltonian up to first order in $\VV{SO}$ can be given for the $N_0+1$ particle subblock:
\begin{align}\label{H_eff_ani}
\operatorname{H}_{0}^{N_0+1} = E_{N_0+1}^\mathrm{g} - J^\mathrm{ex}_{SL}\left( \hat{S}^2-1 \right)  + \lambda_1\,\hat\tau\hat S_z.
\end{align}
Equation~\eqref{H_eff_ani} is one major result of this work. It shows that, similar to the well studied molecular magnets~\cite{Mannini2010,Chiesa2013,Wegewijs2015,Burzuri2015},
the interplay of spin-orbit coupling and exchange interactions yield magnetic anisotropies which can be captured by effective spin Hamiltonians. Noticeably, because Eq.~\eqref{H_eff_ani}
was derived from the microscopic molecular Hamiltonian $\HH{mol}$, it was possible to check
that deviations are in the $\mu$eV range and only of quantitative nature by comparison of the spectrum to the numerically evaluated one.

\subsection{Interaction with magnetic fields}

\begin{figure}
 \includegraphics[width=\columnwidth]{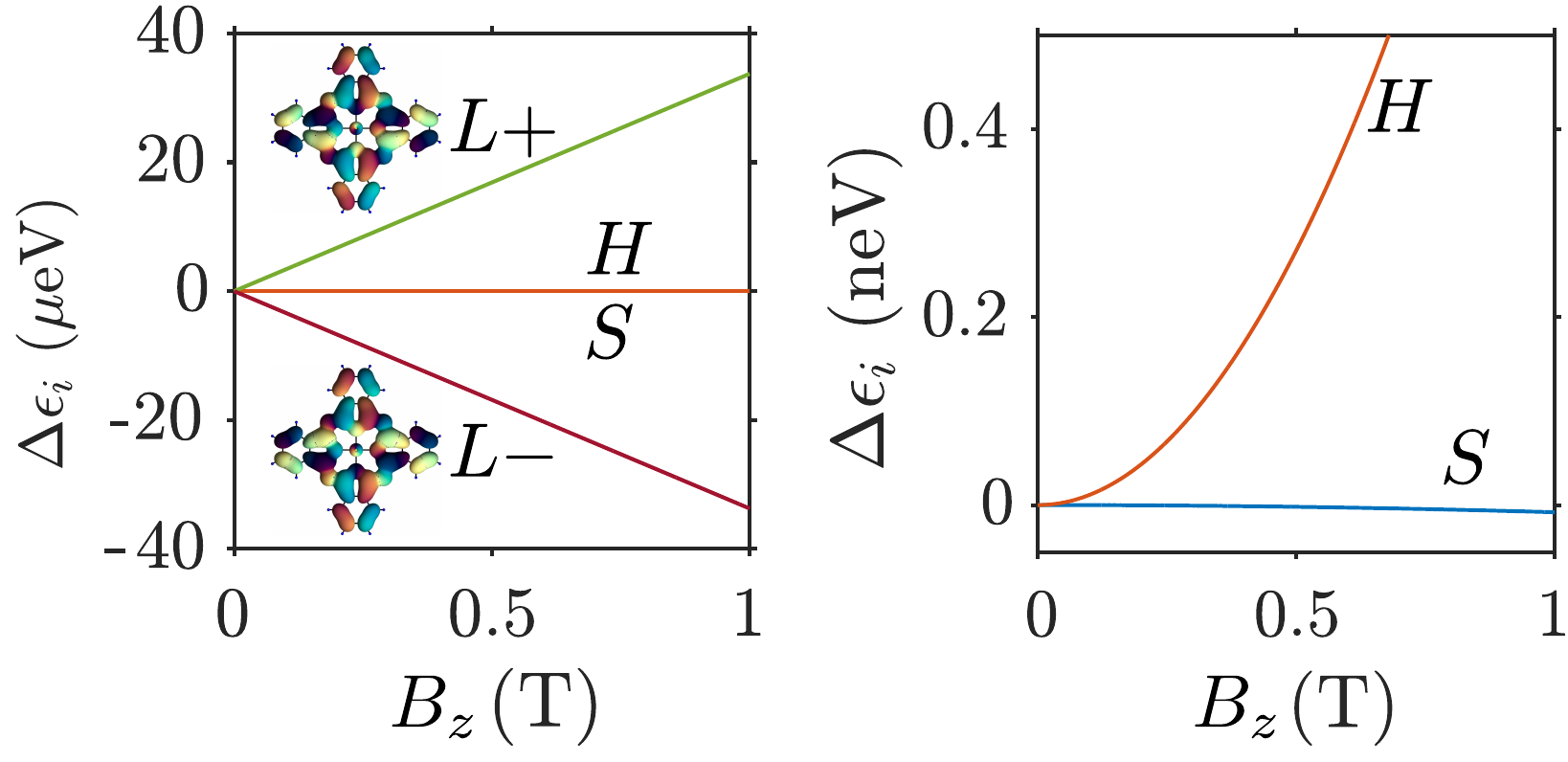}
 \caption{(a) Dependence of the single particle orbital energies on the magnetic field strength. From this, the effective orbital moment of the LUMOs, here depicted in their complex representation,
 can be extracted as $\mu_\mathrm{orb}=33.7~\mu$eVT$^{-1}$. 
 The energies of the SOMO and HOMO orbitals depend quadratically on the magnetic field and involve a much lower scale than the LUMOs, as seen in the close-up in panel (b).}\label{fig_3}
\end{figure} 

An experimentally accessible way to probe magnetic anisotropies is to apply external magnetic fields.
In order to account for interactions of orbitals with magnetic fields, the atomic hopping matrix elements $b_{\alpha\beta}$ in Eq.~\eqref{Eq:H_onebody} have to be corrected with Peierls phase factors,
\begin{align}
  b_{\alpha\beta} \rightarrow b_{\alpha\beta}\,\ee{i\phi_{\alpha\beta}},
\end{align}
where, using the gauge $\mathbf{A}=-B_z\,y\hat{x}$, the phase is given by
\begin{align}
 \phi_{\alpha\beta} = \frac{eB_z}{2\hbar}\left( y_\alpha + y_\beta \right)\left( x_\alpha - x_\beta \right).
\end{align}
Here $(x_\alpha,y_\alpha)$ are the in-plane atomic coordinates. 
Owing to the planar geometry of CuPc, $\phi_{\alpha\beta}$ depends only on the $z$-component $B_z$ of the magnetic field $\mathbf{B}$.
In Fig.~\ref{fig_3} we show the dependence of the energies of the frontier molecular orbitals on the strength of the magnetic field in $z$-direction, $B_z$. For the two LUMOs we observe
a linear dependence on the magnetic field, yielding an effective orbital moment of $\mu_\mathrm{orb}=33.7~\mu$eVT$^{-1}$. Hereby the LUMO$-$($+$) goes down (up) in energy 
with $B_z$, see Fig.~\ref{fig_3} (a).
The energies of the HOMO and the SOMO however scale quadratically with the magnetic field at a much lower scale, cf. Fig.~\ref{fig_3} (b). This behaviour is expected, since the $a_{1u}$ and $b_{1g}$ representations
have characters $+1$ under $C_2'$ rotations, which transform $B_z$ to $-B_z$. Thus the energies of HOMO and SOMO can not depend on the sign of $B_z$ and must move at least quadratically with $B_z$. The 
two-dimensional $e_g$ representation on the other hand has zero character under $C_2'$ rotations, which implies that the constituents of $e_g$ transform under such rotations either with different signs
or into each other; indeed under a $C_2'$ rotation LUMO$+$ is mapped onto LUMO$-$ and vice versa.

Finally, the interaction of electronic spins with magnetic fields is represented by adding a Zeeman term $\VV{Z}$ to Eq.~\eqref{Eq:H_mol},
\begin{align}
 \HH{mol} \rightarrow \HH{mol} + \VV{Z} = \HH{mol} + g_S\mu_\mathrm{B}\,\hat{\mathbf{S}}\cdot\mathbf{B},
\end{align}
where $g_S=2$ and $\mathbf{S}$ is the total spin operator on the molecule written in the frontier orbital basis.

\subsubsection{Effective low-energy Hamiltonian}

Putting everything together, an effective low-energy Hamiltonian including magnetic interaction terms for both orbital and spin
degrees of freedom can thus be given. It reads
%
\begin{align}\label{Eq:effective_formula}
 \operatorname{H}_\mathrm{eff}^{N} = \operatorname{H}_{0}^N + \mu_\mathrm{orb}\,\hat\tau B_z + g_S\mu_\mathrm{B}\,\hat{\mathbf{S}}\cdot\mathbf{B},  
\end{align}
where $\operatorname{H}_{0}^N $ is the Hamiltonian for the corresponding low-energy $N$-particle subblock as given by Eqs.~\eqref{H_eff_neutral} and~\eqref{H_eff_ani}. 

\subsection{Dynamics and transport}

\subsubsection{Reduced density operator and current}

The transport calculations for the molecule in an STM setup are done by using the formalism introduced in earlier works~\cite{benzene_sandra,CuPc_interference,h2pcpaper}.
For the sake of clarity, in the following we briefly discuss the main steps to obtain the current through the molecule.
The full system is described by the Hamiltonian 
\begin{align}\label{H_full}
 \HH{} = \HH{mol} + \HH{ic} + \HH{S} + \HH{T} + \HH{tun},
\end{align}
where $\HH{mol}$ describes the isolated molecule, see Eq.~\eqref{Eq:H_mol}. To incorporate image charge effects in our model, leading to
renormalizations of the energies of the systems charged states~\cite{image_charges}, we included a term $\HH{ic}$~\cite{sco_cupc_arxiv},
\begin{align}
 \HH{ic} = -\delta_\mathrm{ic}\left( \hat N - N_0 \right)^2,
\end{align}
where $\hat N$ is the particle number operator on the molecule.
Electrostatic considerations regarding the geometry of the STM setup yielded $\delta_\mathrm{ic}\approx0.3~$eV~\cite{sco_cupc_arxiv}.  
The Hamiltonians $\HH{S}$ and $\HH{T}$ corresponding to substrate (S) and tip (T), respectively, are describing noninteracting electronic leads. They read
\begin{align}
 \hat{\operatorname{H}}_{\eta=\mathrm{S,T}}=\sum_{\kk\sigma}\eps{\eta\kk}\,\erz{c}{\eta\kk\sigma}\ver{c}{\eta\kk\sigma},
\end{align}
 where $\erz{c}{\eta\kk\sigma}$ creates an electron in lead $\eta$ with spin $\sigma$ and momentum $\kk$.
The tunneling Hamiltonian $\HH{tun}$ finally is given by
\begin{align}
 \HH{tun} = \sum_{\eta\kk i\sigma} t^\eta_{\kk i}\, \erz{c}{\eta\kk\sigma}\ver{d}{i\sigma} +\mathrm{h.c.}.
\end{align}
It contains the tunneling matrix elements $t^\eta_{\kk i}$, which are obtained by calculating
the overlap between the lead wavefunctions $\ket{\eta\kk}$ and the molecular orbitals $\ket{i}$~\cite{benzene_sandra}.

\begin{figure}
 \includegraphics[width=\columnwidth]{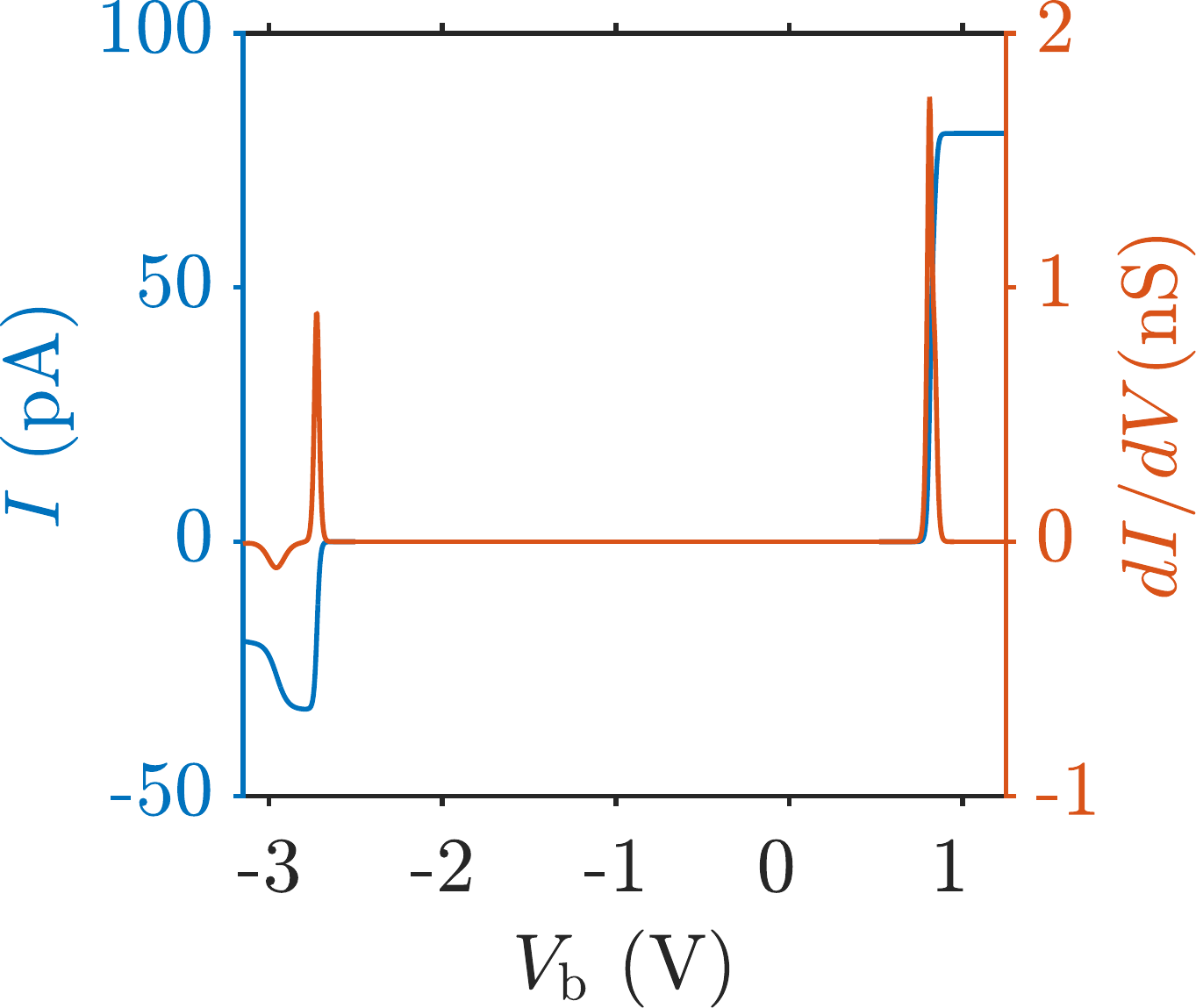}
 \caption{Current and differential conductance curves exhibiting the anionic (cationic) resonance at positive (negative) bias voltage. 
 Note that in contrast to all other results in this work, this curve is taken at a temperature of $60~$K to emphasize the resonances in the $dI/dV$ curve.}\label{fig_4}
\end{figure}

Finally, the dynamics of the transport itself is calculated by evaluating the generalized master equation,
\begin{align}\label{GME}
 \dot{\rho}_\mathrm{red} = \mathcal{L}[\rho_\mathrm{red}],
\end{align}
for the reduced density operator~\cite{benzene_dana_georg,benzene_sandra} 
$\rho_\mathrm{red}=\operatorname{Tr}_{\mathrm{S,T}}\left(\rho\right)$.
The Liouvillian superoperator
\begin{align}
 \mathcal{L} = \mathcal{L}_\mathrm{S} + \mathcal{L}_\mathrm{T} + \mathcal{L}_\mathrm{rel}
\end{align}
contains the terms $\mathcal{L}_\mathrm{S}$ and $\mathcal{L}_\mathrm{T}$ describing tunneling from and to the substrate and the tip, respectively.
To account for relaxation processes leading to de-excitation of molecular excited states,
we included a relaxation term $\mathcal{L}_{\mathrm{rel}}$, analogously to Ref.~\cite{koch_oppen}:
\begin{align}\label{L_rel}
  \mathcal{L}_\mathrm{rel}\left[\rho\right] = -\frac{1}{\tau}\left( \rho - \rho^{\mathrm{th},N}_{kk}\ket{Nk}\bra{Nk}\sum_l \rho^N_{ll} \right).
\end{align}
%
It depends on the deviation of $\rho$
from the thermal solution $\rho^\mathrm{th}$, which is given by a Boltzmann distribution:
\begin{align}
 \rho^\mathrm{th}=\sum_{Nk} \frac{ \ee{-\beta E_{Nk} } }{ \sum_l \ee{-\beta E_{Nl} } } \ket{Nk}\bra{Nk},
\end{align}
 with $\beta=\left(k_\mathrm{B}T\right)^{-1}$.
Since $\mathcal{L}_\mathrm{rel}$ acts separately on each $N$-particle subblock, it conserves the particle number on the molecule and thus does not contribute to transport directly.
In this work, the relaxation factor $\frac{1}{\tau}$ is around the same order of magnitude as the mean tip tunneling rate onto the molecule. 
In particular, we are interested in the stationary solution $\rho_\mathrm{red}^\infty$ for which $\dot\rho_\mathrm{red}^\infty=\mathcal{L}[\rho_\mathrm{red}^\infty]=0$.
Finally, the current through the system in the stationary limit can be evaluated as
\begin{align}\label{Eq_current}
\braket{\hat I_\mathrm{S}+\hat I_\mathrm{T}}=\frac{\mathrm{d}}{\mathrm{d}t}\braket{\hat N}=\operatorname{Tr}_\mathrm{mol}\left(\hat N\mathcal{L}[\rho_\mathrm{red}^\infty]\right)=0,
\end{align}
yielding the current operator for lead $\eta$ as $\hat I_\eta=\hat N\mathcal{L_\eta}$.

\begin{figure}
 \includegraphics[width=\columnwidth]{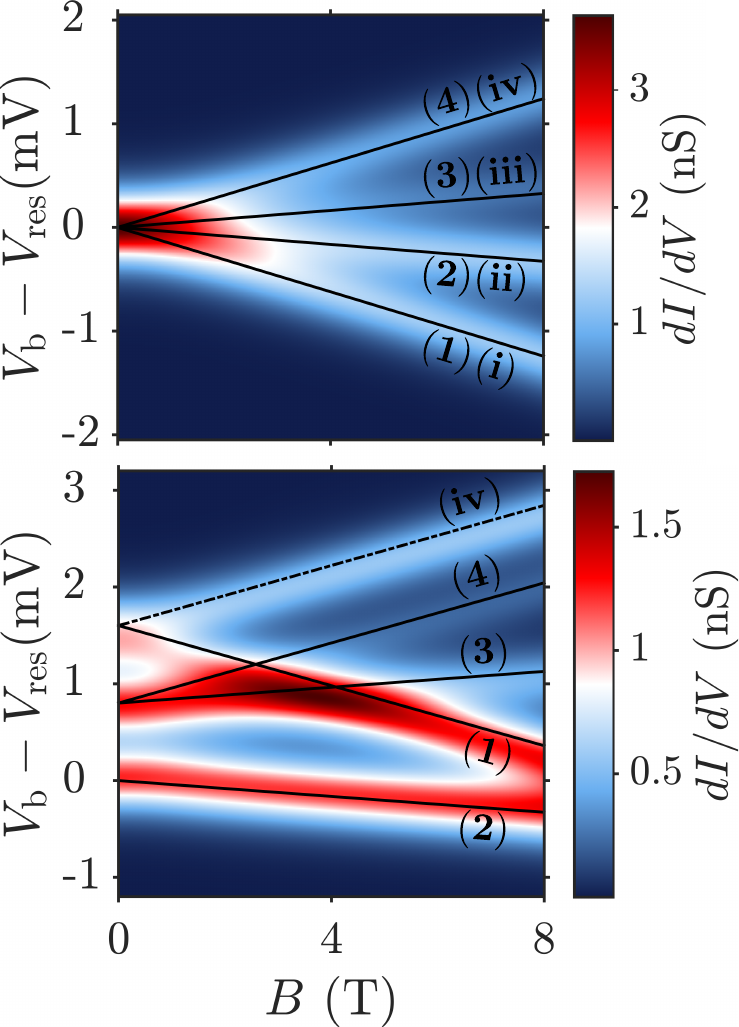}
 \caption{Differential conductance maps as a function of the strength $B_z$ of the magnetic field in $z$-direction. Upper (lower) panel: Spin-orbit interaction
 switched off (on). Solid and dashed lines depict the addition spectrum as calculated from the effective spin Hamiltonian, cf. Eq.~\eqref{Eq:effective_formula}. 
 Transitions starting from the neutral groundstate are denoted by solid lines, those from the neutral excited state by dashed lines.}\label{fig_5}
\end{figure}

\subsubsection{Transport characteristics}
 
In this work, a tip-molecule distance of 5~\AA~was used and simulations were done at the temperature $T=1~$K. 
We assumed a substrate workfunction of $\phi_0=4.65~$eV and a renormalization of the single particle energies
$\delta_i=\delta=1.5~$eV (cf. Eq.~\eqref{Eq:renorm}). 
Numerical results for the current and the differential conductance, according to Eq.~\eqref{Eq_current} and using the full Hamiltonian $\HH{mol}$ in Eq.~\eqref{H_full}, are shown in Fig.~\eqref{fig_4}.
Anionic (cationic) resonances at positive (negative) bias voltages are clearly seen.
We find a very good agreement between our numerically evaluated positions
of the cationic and anionic resonances with those of the experiment in Ref.~\cite{repp_charge_state}, where a Cu(100) substrate was used.

Notice that, in our model, the bias voltage at which a tip-mediated transition from the $m$th neutral state to the $n$th anionic state of the molecule is happening is
\begin{align}\label{V_res}
 V_{\mathrm{res},mn} = \frac{1}{\alpha_\mathrm{T}|e|}\left( E_{N_0+1,n} - E_{N_0,m} - \delta_\mathrm{ic} + \phi_0 \right),
\end{align}
where $e$ is the electron charge and $\alpha_\mathrm{T}$ accounts for the fact that in STM setups the bias voltage drops asymetrically across the junction.
We are using $\alpha_\mathrm{T}=0.59$ for the tip and $\alpha_\mathrm{S}=-0.16$ for the substrate~\cite{sco_cupc_arxiv}.
If given without indices, $V_\mathrm{res}$ denotes the bias voltage corresponding to the groundstate-to-groundstate resonance.

The negative differential conductance at large negative bias in Fig.~\ref{fig_4} is caused by blocking due to population of excited states of the molecule.
This has already been discussed in some previous work~\cite{CuPc_interference} and will not be of further interest here.

\begin{figure*}
  \includegraphics[width=\textwidth]{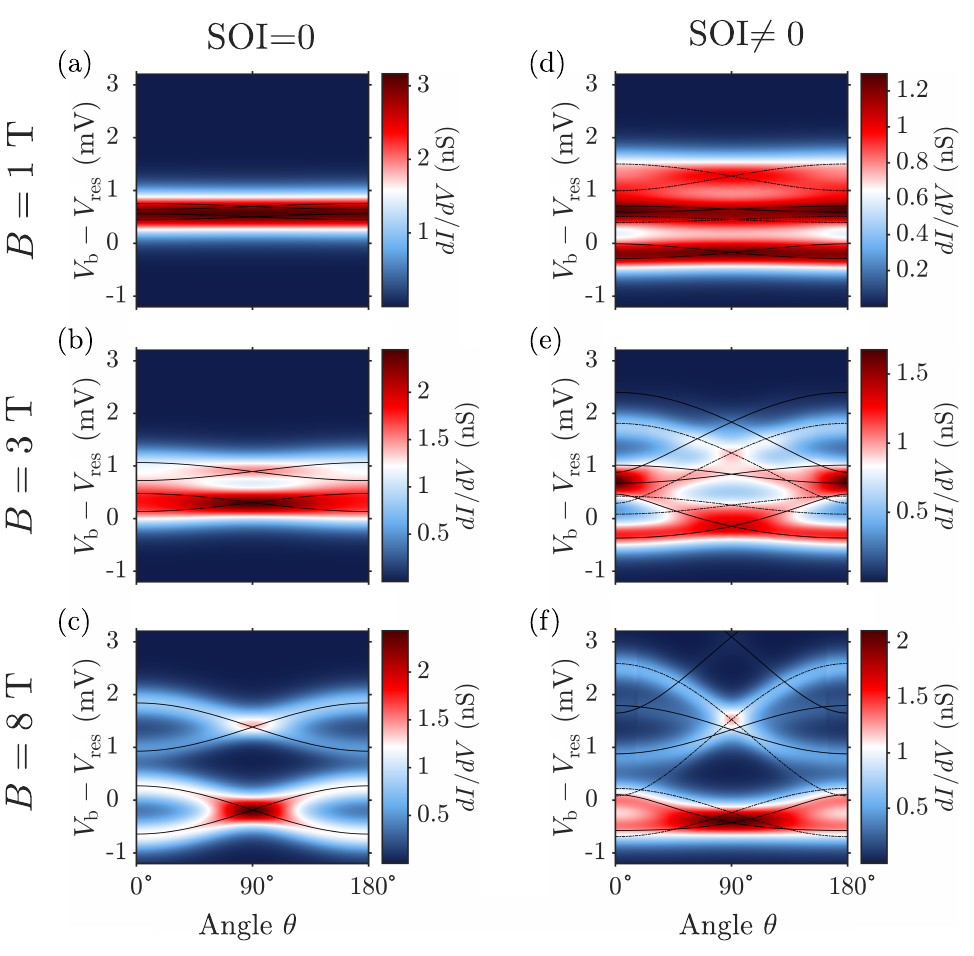}
  \caption{Differential conductance maps vs. the angle $\theta$, formed by the applied magnetic field with the $z$-axis. Left (right) panels are without (with) SOI.
  Upper, middle and lower panels are calculated for a magnetic field strength of $1~$T, $3~$T and $8~$T, respectively.
  Solid and dashed lines depict the addition spectrum as calculated from the effective spin Hamiltonian, cf. Eq.~\eqref{Eq:effective_formula}. 
 Transitions starting from the neutral groundstate are denoted by solid lines, those from the neutral excited state by dashed lines.}\label{fig_6}
 \end{figure*}

\subsubsection{Transport simulations at finite magnetic fields}

In Fig.~\ref{fig_5} we show the splitting of the anionic resonance with applied magnetic field in a $dI/dV$ map. 
In the upper panel SOI is switched off, whereas in the lower panel it is switched on. One striking difference at
first glance is the zero-field splitting for nonvanishing SOI, which is proportional to $\lambda_1$ but enhanced by the bias drop, cf. Eq.~\eqref{V_res}.
For vanishing SOI, when $S_z$ is a good quantum number, we can readily identify the corresponding transitions
by using the effective spin Hamiltonian introduced in Eq.~\eqref{Eq:effective_formula}.
In the following, transitions from the neutral groundstate will be denoted by arabic numbers:
\begin{align}
 (1):\quad		\ket{N_0,\down} &\rightarrow \ket{\mathbf{T}_-^-}\phantom{,}		\nonumber\\
 (2):\quad		\ket{N_0,\down} &\rightarrow \ket{\mathbf{T}_+^-}\phantom{,}		\nonumber\\
 (3):\quad		\ket{N_0,\down} &\rightarrow \ket{\mathbf{T}_-^0}\phantom{,}		\nonumber\\
 (4):\quad		\ket{N_0,\down} &\rightarrow \ket{\mathbf{T}_+^0},			\nonumber
\end{align}
while transitions from the neutral excited state will be denoted by Roman numerals:
\begin{align}
 (i):\quad		\ket{N_0,\up} &\rightarrow \ket{\mathbf{T}_-^0}\phantom{,}		\nonumber\\
 (ii):\quad		\ket{N_0,\up} &\rightarrow \ket{\mathbf{T}_+^0}\phantom{,}		\nonumber\\
 (iii):\quad		\ket{N_0,\up} &\rightarrow \ket{\mathbf{T}_-^+}\phantom{,}		\nonumber\\
 (iv):\quad		\ket{N_0,\up} &\rightarrow \ket{\mathbf{T}_+^+}.			\nonumber
\end{align}
Other transitions are forbidden due to the selection rule for $S_z$, $\Delta S_z=\pm\frac{1}{2}$.
The reason for the splitting into four lines observed in Fig.~\ref{fig_5} top is that the orbital moment of the LUMO is not of the same size as the Bohr magneton. 
 
For nonvanishing SOI, see lower panel of Fig.~\ref{fig_5}, the definite assignment of transitions is not straightforward, at least for small magnetic fields.
Since $\mathbf T_+^-$ and $\mathbf T_-^+$ are shifted downward by SOI, transition $(2)$ now is the lowest lying transition,
whereas transition $(1)$ is shifted upward due to the positive contribution $+\lambda_1$ to $\mathbf T_-^-$.
Furthermore, transition $(iv)$ is the only excited state transition which can be definitely assigned to a line in the lower panel in Fig.~\ref{fig_5}.

Figure~\ref{fig_6} finally shows $dI/dV$ maps as a function of the angle $\theta$ between the magnetic field and the $z$-axis. Hereby panels (a), (b) and (c)
show results obtained with vanishing SOI and panels (d), (e) and (f) are for finite SOI.
Again, the results were fitted using the effective spin Hamiltonian introduced in Eq.~\eqref{Eq:effective_formula} with good agreement.
The respective transitions can be identified by checking the assigned transitions in Fig.~\ref{fig_5} at the corresponding field strength.

Already at $|\mathbf{B}|=B=1~$T, cf. (a) and (d),
the influence of SOI can be clearly seen. While for vanishing SOI any anisotropy of the $dI/dV$ map is hidden beneath the temperature broadening,
for finite SOI a slight $\theta$-dependence can be observed. For $B=3~$T, now also in the vanishing SOI case, Fig.~\ref{fig_6} (b), a slight anisotropy
due to the orbital moment of the LUMOs can be observed, although still blurred by temperature. Again, at finite SOI in Fig.~\ref{fig_6} (e) there is a much more pronounced
dependence on $\theta$. The high conductance areas at $\theta=0^\circ$ and $\theta=180^\circ$ for $V_\mathrm{b}-V_\mathrm{res}\approx0.8~$meV correspond to 
the high conductance area in the middle of Fig.~\ref{fig_5} bottom,
where many transitions are taking place at the same time. At $B=8~$T, the magnetic field is dominating and a characteristic double cosine-like behaviour of the resonances
can be observed, for both the case with no SOI, Fig.~\ref{fig_6} (c), and finite SOI, Fig.~\ref{fig_6} (f). For vanishing SOI, this behaviour is caused by the orbital moment of the LUMOs, 
since they interchange their positions when going from $B_z$ to $-B_z$. The  overall splitting between the double cosines, most evident at $\theta=90^\circ$, is caused by the Zeeman term.
The results for $B=8~$T in Fig.~\ref{fig_6} (f) at finite SOI are similar to those in Fig.~\ref{fig_6} (c),
with the only difference that the cosine at large biases is more stretched, the one at low bias more compressed.

\section{Conclusions}\label{Sec3}
We established a model Hamiltonian for CuPc which accounts for electron-electron, spin-orbit and magnetic interactions in a minimal single particle basis
represented by four frontier orbitals; the SOMO, the HOMO and two degenerate LUMOs. 
The distinct properties of these orbitals under rotations allowed us to deduce selection rules for matrix elements
of the Coulomb interaction, which drastically reduce the number of nonvanishing terms and simplify the numerical diagonalization of the full many-body Hamiltonian.
For the low-energy parts of the neutral and anionic blocks of the many-body spectrum we could further derive an effective spin Hamiltonian, 
capturing both SOI induced splittings and magnetic anisotropy. In order to study fingerprints of the SOI under realistic experimental conditions, 
we have studied the magnetotransport characteristics of a CuPc based junction in an STM setup. 
To this extent, a generalized master equation for the reduced density matrix associated to the full many-body Hamiltonian had to be solved 
in order to numerically obtain both the current and the differential conductance. 
Noticeably, by using the effective spin Hamiltonian, it was possible to reconstruct the nature of the many-body 
resonances observed  in the numerical calculations.

In summary,  we believe that our work significantly advances the present understanding of spin properties of CuPc; 
moreover, the flexibility of our model Hamiltonian approach opens new perspectives for the investigation of other configurationally  similar  metallorganic compounds.

\acknowledgments{The authors thank Thomas Niehaus, Jascha Repp and Dmitry Ryndyk for fruitful discussions.
 Financial support by the Deutsche Forschungsgemeinschaft within the research program SFB 689 is acknowledged.}

\appendix

\section{Transformation from the atomic to the molecular orbital basis}\label{App:AO2MO}
The Hamiltonian of a molecule in Born-Oppenheimer approximation, after dismissing terms which only depend on the positions of the nuclei and are
therefore constant, can be written as
 \begin{align}\label{Eq:general}
  \HH{} = &\sum_{\substack{\alpha\beta\sigma\\mn}}\left(\  h_{\alpha m,\beta n} +  \Delta V^\mathrm{ion}_{\alpha m,\beta n} \right)\,%
  \erz{d}{\alpha m\sigma}\ver{d}{\beta n\sigma} \nonumber\\
	+\frac{1}{2} &\sum_{\substack{\alpha\beta\gamma\delta\\mnop}}\sum_{\sigma\sigma'} %
   V_{\alpha\beta\gamma\delta}^{mnop}\,
  \erz{d}{\alpha m\sigma}\erz{d}{\gamma p\sigma'}\ver{d}{\delta q\sigma'}\ver{d}{\beta n\sigma},
 \end{align}
 where $\erz{d}{\alpha m\sigma}$ creates an electron in the atomic orbital $\ket{\alpha m\sigma}$ with 
 orbital quantum number $m$ and spin $\sigma$ centered at atom $\alpha$.
 Further we have defined
 \begin{align}
 \ h_{\alpha m,\beta n} &:= \eps{\alpha m}\delta_{\alpha\beta}\delta_{mn} + b_{\alpha m,\beta n},
 \end{align}
 where $\eps{\alpha m}$ is the energy of orbital $m$ on atom $\alpha$
 and $b_{\alpha m,\beta n}$ is the hopping integral between orbital $m$ on atom $\alpha$ and orbital $n$ on atom $\beta$.
 All non-hopping terms can be condensed in the crystal field correction
 \begin{align}\label{crystal}
  \Delta V^\mathrm{ion}_{\alpha m,\beta n} &:= \sum_{\gamma}^{\gamma\neq\alpha,\beta}\braket{\alpha m\sigma|\VV{\gamma}|\beta n \sigma},
 \end{align}
 where $\hat V_\gamma$ is the atomic core potential at $\rr_\gamma$. Equation~\eqref{crystal} defines the crystal field correction to the single particle Hamiltonian.
 Finally, we have the ordinary matrix elements $V_{\alpha\beta\gamma\delta}^{mnop}$ of the Coulomb interaction.
 
 The $h_{\alpha m,\beta n}$ are elements of a matrix $\mathbf{h}$ which corresponds to the single particle Hamiltonian of the molecule with only onsite energies and hopping terms.  
 After performing a transformation to the molecular orbital basis, in which $\mathbf{h}$ is diagonal, $\ket{i\sigma}    = \sum_{\alpha m} c_{i\alpha m}\, \ket{\alpha m\sigma}$,
 and using the approximation that the basis $\ket{\alpha m\sigma}$ is orthogonal, the Hamiltonian reads:
 \begin{align}\label{H_appendix}
  \HH{} &= \sum_{ij\sigma}\left(\eps{i}\delta_{ij} + \Delta V^\mathrm{ion}_{ij} \right)\, \erz{d}{i\sigma}\ver{d}{j\sigma} \nonumber \\
  +\frac{1}{2}  & \sum_{ijkl}\sum_{\sigma\sigma'}V_{ijkl}\,\erz{d}{i\sigma}\erz{d}{k\sigma'}\ver{d}{l\sigma'}\ver{d}{j\sigma},
 \end{align}
 where $\Delta V^\mathrm{ion}_{ij}$ now is
 \begin{align}\label{crystal_MO}
 \Delta V^\mathrm{ion}_{ij} &= \sum_{\substack{\alpha\beta\\ mn}}c_{i\alpha m}^*c_{j\beta n}\,\Delta V^\mathrm{ion}_{\alpha m,\beta n}.
 \end{align}

\section{Symmetries in the frontier orbitals basis}\label{Sec:reduction}

\begin{table}[ht!]
  \begin{ruledtabular}
   \begin{tabular}{llll}
    $U_S$						&	11.352 eV		&	$J^\mathrm{ex}_{HL}=-\tilde J^\mathrm{p}_{HL+L-}$&	548 meV		\\
    $U_H$						&	1.752 eV		&	$J^\mathrm{ex}_{L+L-}$ 				&	258 meV		\\
    $U_L=U_{L+L-}$					&	1.808 eV		&	$J^\mathrm{p}_{L+L-}$				&	168 meV		\\
    $U_{SH}$						&	1.777 eV		& 	$J^\mathrm{ex}_{SL}=-\tilde J^\mathrm{p}_{SL+L-}$&	9  meV 		\\
    $U_{SL}$						&	1.993 eV 		&	 $J^\mathrm{ex}_{SH}=J^\mathrm{p}_{SH}$	&		2 meV 		\\
    $U_{HL}$						&	1.758 eV		& 			&			\\
   \end{tabular}
  \end{ruledtabular}
  \caption{Major nonvanishing Coulomb integrals between the SOMO($S$), the HOMO($H$), the LUMO$+$ and the LUMO$-$. When the LUMOs need to be distinguished, they are denoted as $L+$ or $L-$,
  otherwise just by $L$. All values are calculated numerically using Monte Carlo integration~\cite{gsl} of the real space orbitals depicted in Figs.~\ref{fig_1}(c) and~\ref{fig_3}, respectively,
  and renormalized by a constant $\eps{r}=2.2$.\label{tab_1}}
 \end{table}

 One huge simplification which is possible in the molecular orbital basis is the reduction of the size of our Hilbert space $\mathcal{H}$,
 which occurs by retaining few relevant molecular orbitals only.
 To this end we split the full molecular basis into frozen and dynamic orbitals, where $N_f$ of the frozen orbitals are assumed to be always fully occupied and the remaining $N_e$ set to be always empty. 
We do not make any assumption about the occupation of the $N_d$ dynamic states. Whether these $N_d$ frontier orbitals are full or empty depends on the electrochemical potential of
the molecule, and on whether an exchange of electrons with the environment is possible.
 
 In the occupation number representation a general state of the Fock space then looks like
 \begin{align}\label{frozenket}
  \ket{\Psi} \approx 		\underbrace{{\ket{11\ldots11}}}_{2N_f} \otimes%
				\underbrace{\ket{n_{k\uparrow}n_{k\downarrow}\ldots n_{l\uparrow}n_{l\downarrow}}}_{2N_d} \otimes %
				\underbrace{\ket{00\ldots00}}_{2N_e}.	
 \end{align} 
In this work we assume the molecule to be neutral under equilibrium conditions, with 195 valence electrons.
Thus, the orbitals we choose to build up the subspace of dynamic orbitals are orbitals Nrs. 97-100, see Fig.~\ref{fig_1} (b).
 This choice results in the lowest 96 molecular orbitals being doubly filled.
 Note that the choice of the LUMO states $L\pm$ rather than $L_{zx/yz}$ is convenient due to the fact that these orbitals
 acquire a definite phase upon rotations of 90 degrees around the main symmetry axis of the molecule.
 Specifically, for the four frontier orbitals $S$, $H$ and $L\pm$, the acquired phases are $\phi_S=\pi$, $\phi_H=0$ and $\phi_{L\pm}=\pm\frac{\pi}{2}$, respectively.
 This in turn imposes symmetry constraints on the Hamiltonian~\eqref{H_appendix}. Consider e.g. the Coulomb interaction
 \begin{widetext}
  \begin{align}
 V_{ijkl} = \frac{1}{4\pi\varepsilon_0}\iint\mathrm{d}^3r_1\,\mathrm{d}^3r_2\,\psi_i^*(\rr_1)\psi_j(\rr_1)\frac{1}{|\rr_1-\rr_2|}\psi_k^*(\rr_2)\psi_l(\rr_2).
\end{align}
 \end{widetext}
%
 Then, in the frontier orbital basis it holds that:
 \begin{align}
  V_{ijkl} &= \mathrm{e}^{-i(\phi_i-\phi_j+\phi_k-\phi_l)}\,V_{ijkl}.
 \end{align}
Therefore a given matrix element of the Coulomb interaction $V_{ijkl}$ is different from zero only if the sum of the corresponding phases adds up
to multiples of $2\pi$: $\phi_i-\phi_j+\phi_k-\phi_l=2\pi\cdot n,~ n\in\mathbb{Z}$. 
In Tab.~\ref{tab_1} we list all nonvanishing matrix elements of the Coulomb interaction which are
used in this work. For the crystal field correction $\Delta V^\mathrm{ion}_{ij}$ it can be shown that:
 \begin{align}
 \Delta V^\mathrm{ion}_{ij} &= \mathrm{e}^{-i(\phi_i-\phi_j)}\,\Delta V^\mathrm{ion}_{ij} \\
 \Rightarrow \Delta V^\mathrm{ion}_{ij} &= \Delta V^\mathrm{ion}_{ii}\,\delta_{ij},
 \end{align}
 since all phases $\phi_i$ are different; $\phi_i\neq\phi_j$ for $i\neq j$. 
 Hence $\Delta V^\mathrm{ion}_{ij}$ is diagonal in the \{$S$,$H$,$L\pm$\} basis. In the following we treat the
 $\Delta V^\mathrm{ion}_{ii}$ as free parameters and include them in the paramteter $\delta_i$ entering Eq.~\eqref{Eq:renorm}.
 
 \section{Details on the perturbative treatment of SOI}\label{App:SO1}\label{App:SO2}

In addition to the states introduced in Eq.~\eqref{main_states}, the following states must be also taken into account when
performing second order perturbation theory:
\begin{align}\label{add_states}
	\ket{L\tau\up,L\tau\down} &= \erz{d}{L\tau\up}\erz{d}{L\tau\down}\ket{\Omega},	\nonumber\\
	\ket{L\tau\sigma,L\bar\tau\sigma'} &= \erz{d}{L\tau\sigma}\erz{d}{L\tau\sigma'}\ket{\Omega},	\nonumber\\
	\ket{S\up,S\down} &= \erz{d}{S\up}\erz{d}{S\down}\ket{\Omega},	
\end{align}
with $E_{L\tau\up,L\tau\down}=E_{L\tau\sigma,L\bar\tau\sigma'}=\Delta_1$ and $E_{S\up,S\down} =\Delta_2$. 
In the basis introduced in Eqs.~\eqref{main_states} and \eqref{add_states}, $\VV{SO}$ is blockdiagonal and decomposes into six subblocks: 
two three-dimensional, two two-dimensional, one four-dimensional and one one-dimensional subblocks.
%

%
%
%
%
%
%

The four dimensional subblock describes the effects of SOI on the $\mathbf T_+^+$ and $\mathbf T_-^-$ states.
Written in the basis \{$\ket{\mathbf T_+^+}$,$\ket{\mathbf T_-^-}$,$\ket{L^+\up,L^-\down}$,$\ket{S\up,S\down}$\}, the Hamiltonian reads
\begin{align}
	H=&\begin{pmatrix}
	-J^\mathrm{ex}_{SL}	&	0			&	0		&	0	\\
	0			&	-J^\mathrm{ex}_{SL}	&	0		&	0	\\
	0			&	0			&	\Delta_1	&	0	\\
	0			&	0			&	0		&	\Delta_2
	\end{pmatrix}
\nonumber\\%
+&
	\begin{pmatrix}
	\lambda_1		&	0				&	-\sqrt{2}\lambda_2		&	\sqrt{2}\lambda_2				\\
	0			&	\lambda_1			&	\sqrt{2}\lambda_2		&	-\sqrt{2}\lambda_2				\\
	-\sqrt{2}\lambda_2	&	\sqrt{2}\lambda_2		&	\lambda_1			&	0				\\
	\sqrt{2}\lambda_2	&	-\sqrt{2}\lambda_2		&	0				&	0				
	\end{pmatrix}.
\end{align} 
The degeneracy of the unperturbed states $\mathbf T_+^+$ and $\mathbf T_-^-$ and the fact that there are no matrix-elements 
which couple these states require the use of second order degenerate perturbation theory.
Applying it yields the following matrix $M$:
\begin{align}
	M= A\cdot
		\begin{pmatrix}
			1	&	-1	\\
			-1	&	1
		\end{pmatrix},
\end{align} 
where the prefactor $A$ is given by
\begin{align}
	A = -2\lambda_2^2\left( \frac{1}{\Delta_1+J_{SL}^\mathrm{ex}} + \frac{1}{\Delta_2+J_{SL}^\mathrm{ex}} \right).
\end{align} 
Diagonalization of $M$ gives the second-order energy corrections
\begin{align}
	\Delta E(\alpha) &= \lambda_1, \\
	\Delta E(\beta) &= \lambda_1 - 4 \lambda_2^2\left( \frac{1}{\Delta_1+J_{SL}^\mathrm{ex}} + \frac{1}{\Delta_2+J_{SL}^\mathrm{ex}} \right),
\end{align}
and the correct linear combinations of the states $\mathbf T_+^+$ and $\mathbf T_-^-$:
\begin{align}
	\ket{\alpha} &= \oost\Big( \ket{\mathbf T_+^+} + \ket{\mathbf T_-^-} \Big) \\
	\ket{\beta} &= \oost\Big( \ket{\mathbf T_+^+} - \ket{\mathbf T_-^-} \Big).
\end{align}
Writing $H$ in the basis \{$\ket{\alpha}$,$\ket{\beta}$,$\ket{L^+\up,L^-\down}$,$\ket{S\up,S\down}$\} yields:
\begin{align}
	\tilde H = &\begin{pmatrix}
	-J_{SL}^\mathrm{ex}	&	0			&	0		&	0	\\
	0			&	-J_{SL}^\mathrm{ex}	&	0		&	0	\\
	0			&	0			&	\Delta_1	&	0	\\
	0			&	0			&	0		&	\Delta_2	           	
	           \end{pmatrix}\nonumber\\
+&
	\begin{pmatrix}
	\lambda_1		&	0				&	0				&	0				\\
	0			&	\lambda_1			&	-2\lambda_2			&	2\lambda_2				\\
	0			&	-2\lambda_2			&	\lambda_1				&	0				\\
	0			&	2\lambda_2			&	0				&	0				
	\end{pmatrix}.
\end{align} 
We see that $\ket{\alpha}$ stays unaffected by the perturbation, whereas $\ket{\beta}$ will change:
\begin{align}
	\ket{\beta}	\rightarrow	\ket{\beta}	 +& 2\, \frac{\lambda_2}{\Delta_1+J_{SL}^\mathrm{ex}} \ket{L^+\up,L^-\down}  \nonumber\\
							-&  2\, \frac{\lambda_2}{\Delta_2+J_{SL}^\mathrm{ex}} \ket{S\up,S\down}.
\end{align}


The mixing of $\mathbf{T}_+^-$ and $\mathbf{T}_-^+$ is caused by a pair-hopping term in the Hamiltonian, more precisely by
\begin{align}
 \frac{1}{2}J_{L+L-}^\mathrm{p} \sum_\sigma\left( \erz{d}{L+\sigma}\erz{d}{L+\bar\sigma}\ver{d}{L-\bar\sigma}\ver{d}{L-\sigma} + \mathrm{h.c.} \right),
\end{align}
which couples $\mathbf{T}_+^-$ and $\mathbf{T}_-^+$ to the following states:
\begin{align}
 \ket{a}	&= \frac{1}{\sqrt{2}} \erz{d}{H\up} \erz{d}{H\down} \left( \erz{d}{L+\up} \erz{d}{L+\down} -  \erz{d}{L-\up} \erz{d}{L-\down} \right)\ket{0},\nonumber\\
 \ket{b}	&= \frac{1}{\sqrt{2}} \erz{d}{H\up} \erz{d}{H\down} \left( \erz{d}{L+\up} \erz{d}{L+\down} +  \erz{d}{L-\up} \erz{d}{L-\down} \right)\ket{0},
\end{align}
with corresponding energies $E_a$ and $E_b=E_a+2J_{L+L-}^\mathrm{p}$. Then, after introducing
\begin{align}
 \ket{\mathbf{T}_1} &= \frac{1}{\sqrt{2}}\left( \ket{\mathbf{T}_+^-} + \ket{\mathbf{T}_-^+} \right),\nonumber \\
 \ket{\mathbf{T}_2} &= \frac{1}{\sqrt{2}}\left( \ket{\mathbf{T}_+^-} - \ket{\mathbf{T}_-^+} \right),
\end{align}
the Hamiltonian in the basis of these four states can be written as
\begin{align}
 H = \begin{pmatrix*}
      H_{1b} 	& 0\\
      0		& H_{2a}
     \end{pmatrix*},
\end{align}
with
\begin{align}
 H_{1b} = \begin{pmatrix*}
	-J_{SL}^\mathrm{ex}-\lambda_1 	& 	\lambda_2\\
		\lambda_2		& 	E_b 
     \end{pmatrix*}
\end{align}
and 
\begin{align}
  H_{2a} = \begin{pmatrix*}
	-J_{SL}^\mathrm{ex}-\lambda_1 	& 	\lambda_2\\
		\lambda_2		& 	E_a 
     \end{pmatrix*}.
\end{align}
Diagonalization finally yields the four states
\begin{align}
 \ket{1} 	&= \frac{1}{\sqrt{1-\gamma_b^2}} \left( \ket{\mathbf{T}_1} + \gamma_b \ket{b}\right),\nonumber\\
 \ket{2} 	&= \frac{1}{\sqrt{1-\gamma_a^2}} \left( \ket{\mathbf{T}_2} + \gamma_a \ket{a}\right),\nonumber\\
 \ket{\tilde1} 	&= \frac{1}{\sqrt{1-\gamma_b^2}} \left( \ket{b} - \gamma_b\ket{\mathbf{T}_1}\right),\nonumber\\
 \ket{\tilde2} 	&= \frac{1}{\sqrt{1-\gamma_a^2}} \left( \ket{a} - \gamma_a\ket{\mathbf{T}_2}\right),
\end{align}
with the admixture $\gamma_{a/b}\approx\frac{-\lambda_2}{E_{a/b}+J_{SL}^\mathrm{ex}}$.
Their energies are approximately

\begin{align}
 E_{1}		&\approx -\lambda_1 - \frac{\lambda_2^2}{E_b+J_{SL}^\mathrm{ex}+\lambda_1},\nonumber\\
 E_{2}		&\approx -\lambda_1 - \frac{\lambda_2^2}{E_a+J_{SL}^\mathrm{ex}+\lambda_1},\nonumber\\
 E_{\tilde1}	&\approx E_b + \frac{\lambda_2^2}{E_b+J_{SL}^\mathrm{ex}+\lambda_1},\nonumber\\
 E_{\tilde2}	&\approx E_a + \frac{\lambda_2^2}{E_a+J_{SL}^\mathrm{ex}+\lambda_1}.
\end{align}
This analysis reproduces mixing and energy splittings consistent with our numerical calculations.

\bibliography{bib_arxiv}

\end{document}